\documentclass[twocolumn,aps,showpacs,prb]{revtex4}

\usepackage{graphicx}

\begin{document}

\title{Probing the Hofstadter butterfly with the quantum oscillation of magnetization}

\author{W. H. Xu$^{1}$, L. P. Yang$^{1}$}\email{yanglp@itp.ac.cn}
\author{M. P. Qin$^{2}$}
\author{T. Xiang$^{2,1}$}

\affiliation{$^1$Institute of Theoretical Physics, Chinese Academy
of Sciences, P.O. Box 2735, Beijing 100190, China}

\affiliation{$^2$Institute of Physics, Chinese Academy of Sciences,
P.O. Box 603, Beijing, 100190, China}

\date{\today}

\begin{abstract}

We have developed a different quantum transfer matrix method to
accurately determine thermodynamic properties of the Hofstadter
model. This method resolves a technical problem which is intractable
by other methods and makes the calculation of physical quantities of
the Hofstadter model in the thermodynamic limit at finite
temperatures feasible. It is shown that the quantum correction to
the de Haas-van Alphen (dHvA) oscillation of magnetization bears the
energy structure of Hofstadter butterfly. The measurement of this
quantum correction, which can be materialized on the superlattice or
cold atom systems, can reveal unambiguously the Hofstadter fractal
energy spectrum.

\end{abstract}

\maketitle

The Hofstadter butterfly is the fractal phase diagram of crystalline
electrons in a magnetic field\cite{Azbel,Hofstadter}.  This problem
is of particular interest because it is one of the few examples in
physics where the difference between rational and irrational numbers
can be tested by experimental
measurements\cite{Thouless,Albrecht,Langbein}. However, probing the
Hofstadter butterfly is a challenging problem because a tiny change
in the external magnetic field may give rise to radical
reconstructions of the ground state. Some hints of Hofstadter
butterfly were reported in the quantum hall
conductivity\cite{Albrecht}, magnetic transport\cite{Weiss} and
microwave \cite{Kuhl} measurements.

The Hofstadter butterfly results from the interplay between a
uniform magnetic field and a periodic crystal potential of
two-dimensional electron gas. Thirty years ago,
Hofstadter\cite{Hofstadter} computed the energy spectrum of the
Harper equation\cite{harper} and discovered this fractal butterfly
structure as a function of the magnetic flux per lattice cell
$\phi$. The model introduced in his work, now called Hofstadter
model, has since then become a paradigm for quantum systems with
singular continuous spectra and nontrivial topological numbers. This
model has been solved by the Bethe Ansatz\cite{Wiegmann, Hatsugai}
and exact diagonalization\cite{Hasegawa} methods when $\phi$ is
rational $\phi=p/q$ ($p$ and $q$ are mutually prime integers) with
relatively small $q$.

Previous studies of the Hofstadter model have focused on the ground
state. There was also a discussion on the magnetization oscillation
with the chemical potential at zero temperature\cite{Gat}. As the
ground state energy is not an analytic function of applied magnetic
field due to the fractal feature of spectra, the magnetic
susceptibility is not a well defined quantity at zero temperature.
Thermal fluctuation can smear the singularity and remove this
non-analytical feature. However, to study thermodynamic properties
at finite temperatures, especially the lattice correction to the
quantum oscillation of magnetization as a function of magnetic
field, one has to solve this model for arbitrary $\phi$. This is a
very challenging problem since the largest $q$ that can be handled
by the Bethe Ansatz or exact diagonalization is generally less than
1000.

In this Letter, we propose a novel quantum transfer matrix method to
study thermodynamic properties of the Hofstadter model on square
lattices. This method avoids direct diagonalization of the
Hamiltonian and allows the thermodynamic limit to be explored
directly and accurately. In Refs.\onlinecite{Taut}, magnetic quantum
oscillations were obtained by numerical simulations and the magnetic
breakdown approach. We will show that the quantum oscillation of
magnetization is a susceptive physical quantity to probe the
hierarchical structure of the Hofstadter butterfly. The measurement
of the quantum oscillation of magnetization reveal unambiguously the
Hofstadter's fractal energy spectrum.

Let us start by taking the Landau gauge in which the vector
potential to be zero along the $x$-axis. By further taking the plane
wave expansion along the $x$-axis, we then decouple the Hofstadter
model $H$ into a sum of a series of one-dimensional Hamiltonian
$H_k$
\begin{eqnarray} \label{eqn:ham}
H&=&\sum_{k} H_{k},\\
H_{k} & = & \sum_y \Big[ tc^{\dag}_{k,y+1}c_{k,y} +
tc^{\dag}_{k,y}c_{k,y+1}  \nonumber \\
&&  +2t\cos\left(2\pi y\phi-k\right) c^{\dag}_{k,y}c_{k,y} \Big] ,
\end{eqnarray}
where $k=2\pi n/N_x \, (n=0,1,...N_x-1)$ is the momentum of
electrons along the x-axis and $N_x$ is the lattice dimension along
that direction. $y$ is the lattice coordinate of electrons along the
$y$-axis. $\phi$ is the magnetic flux penetrating each plaquette.

Given $k$, the partition function of $H_k$ is defined by
\begin{equation}
Z_{k}=\mathrm{Tr} \exp(-\beta H_{k}),
\end{equation}
where $\beta = 1/ k_BT$ and $T$ is temperature. The partition
function of the whole system is simply a product of $Z_k$ for all
$k$.

To evaluate $Z_k$, let us first divide $H_{k}$ into two parts,
$H_{k}=H_{k,even}+H_{k,odd}$, where
\begin{eqnarray*}
H_{k,odd} & = & \sum_y h_{k, 2y-1},  \\
H_{k,even} & = & \sum_y h_{k,2y},
\end{eqnarray*}
and
\[
h_{k,y}  =  t(c^{\dag}_{k,y+1}c_{k,y} +h.c) +2t\cos\left(2\pi
y\phi-k\right)c^{\dag}_{k,y}c_{k,y} .
\]
The individual terms $h_{k,y}$ in $H_{k,even}$ or $H_{k,odd}$
commute with each other. Thus it is relatively simple to evaluate
thermodynamic quantities of $H_{k,even}$ or $H_{k,odd}$. To utilize
this property, let us divide $\beta$ into $M$ equivalent parts,
$\varepsilon=\beta/M$ and apply the Trotter-Suzuki
formula\cite{trotter, suzuki} to decompose $Z_k$ as
\begin{equation}\label{eqn:trotter}
Z_{k}= \mathrm{Tr}\left( e^{-\varepsilon H_{k,odd} } e^{-\varepsilon
H_{k,even}} \right)^{M}+O(\varepsilon^{2}).
\end{equation}
By inserting completeness identities to the above expression, we
have
\begin{equation}\label{eqn:zk}
Z_{k} = \lim_{\varepsilon \rightarrow 0} \prod_{l=1}^{M}
\prod_{y=1}^{N_y/2} v_{2y-1,2y}^{2l-1,2l} v_{2y,2y+1}^{2l,2l+1},
\end{equation}
where
\begin{equation}\label{eq:v}
v_{y,y+1}^{l,l+1} = \langle n_{y}^{l}n_{y+1}^{l}|e^{-\varepsilon
h_{k,y} }|n_{y}^{l+1}n_{y+1}^{l+1}\rangle.
\end{equation}
The subscripts represent the lattice positions and the superscripts
represent the coordinates in the inverse temperature or Trotter
space.

From Eq.~(\ref{eq:v}), one can define a local transfer operator
$\tau$ whose matrix elements are given by
\begin{equation}
\label{eqn:localtransfer} \tau_{y,y+1}^{l,l+1} = \langle n_{y}^{l},
1-n_{y}^{l+1}|e^{-\varepsilon h_{k,y}} |1 - n_{y+1}^{l},
n_{y+1}^{l+1}\rangle.
\end{equation}
An important step in the calculation below is to define this local
transfer matrix using fermion operators. Through a tedious
calculation, we find that this transfer matrix can be expressed as
an exponent of a quadratic function of fermion operators.
\begin{eqnarray} \label{eqn:exponential}
\tau_{y,y+1}^{l,l+1}&=& u_{k,y} \exp \Big[ p_{k,y} d^{\dag}_{l }
d_{l+1}+ q_{k,y} d^{\dag}_{l+1} d_{l} \nonumber \\
&&  +r_{k,y}\left(d^{\dag}_{l} d_{l}- d^{\dag}_{l+1}
 d_{l+1}\right)\Big].
\end{eqnarray}
where $d$'s are fermion operators defined in the Trotter space and
coefficients $(p_{k,y}, q_{k,y}, r_{k,y})$ are determined by the
following equations
\begin{eqnarray*}
\frac{\sinh s_{k,y}}{s_{k,y}}p_{k,y} &=&-\frac{\gamma_{k,y}
\exp\left(\alpha_{k,y}\right)}{\varepsilon
t\sinh\gamma_{k,y}},\\
\frac{\sinh s_{k,y}}{s_{k,y}} q_{k,y}
&=&-\frac{\gamma_{k,y}\exp\left(-\alpha_{k,y}\right)}{\varepsilon
t\sinh\gamma_{k,y}},\\
\frac{\sinh s_{k,y}}{s_{k,y}} r_{k,y} &=&
-\frac{\alpha_{k,y}}{\varepsilon t} ,
\end{eqnarray*}
$\alpha_{k,y}=-\varepsilon [t\cos\left(2\pi y \phi-
k\right)-\mu/2]$,
$\gamma_{k,y}=\sqrt{\alpha^{2}_{k,y}+\varepsilon^{2}t^{2}}$,
$u_{k,y}=-\varepsilon
t\sinh\gamma_{k,y}\exp(\alpha_{k,y})/\gamma_{k,y}$, and
$s_{ky}=\sqrt{p_{k,y}q_{k,y}+r^{2}_{k,y}}$. $\mu$ is the chemical
potential. In general, for any quadratic Hamiltonian, it can be
shown that the corresponding local transfer matrix can be always
written as an exponent of a quadratic function of fermion
operators\cite{Yang}.

Reversing the order of $l$ and $y$ in Eq.~(\ref{eqn:zk}), one can
then reexpress the partition function as a product of transfer
matrices
\begin{equation} \label{eqn:zk2}
Z_{k}= \lim_{\varepsilon \rightarrow 0} \mathrm{Tr} (
T_{1,2}T_{2,3}\cdots T_{N,1} ),
\end{equation}
where $T_{y,y+1}$ are transfer operators defined by
\begin{eqnarray*}
T_{2y-1,2y}&=&\prod_{l}\tau_{2y-1,2y}^{2l-1,2l}, \\
T_{2y,2y+1}&=&\prod_{l}\tau_{2y,2y+1}^{2l,2l+1}.
\end{eqnarray*}

Since coefficients $(p_{k,y}, q_{k,y}, r_{k,y})$ do not depend on
$l$, the above transfer operators are translationally invariant in
every two unit cells along the Trotter direction. Thus we can block
diagonalize these transfer matrices by taking the Fourier
transformation of fermion operators in the Trotter space. With
further simplification, we find that in the thermodynamic limit
$Z_k$ can be expressed as a product of $N_y$ $2\times 2$ matrices
given by the following formula
\begin{equation}\label{eq:zk2}
Z_{k} = \lim_{\varepsilon \rightarrow 0} \mathrm{Tr} \prod_{\omega}
\prod_y^{N_y/2} \left[ t^-_{k,2y-1}(\omega) t^+_{k,2y}(0) \right] ,
\end{equation}
where $\omega = (2m + 1)\pi/ M$ ($m=1, \cdots, M$) is the imaginary
frequency. $t^\pm_{k,y}(\omega) $ are $2\times 2$ matrices defined
by
\begin{eqnarray}
t^\pm_{k,y}(\omega ) = u_{k,y} \left(
  \begin{array}{cc}
    a_{k,y}^\pm & e^{-i\omega} b_{k,y}^{\pm} \\
    e^{i\omega}b_{k,y}^\mp & a_{k,y}^\mp \\
  \end{array}
\right),
\end{eqnarray}
where
\begin{eqnarray*}
a_{k,y}^{\pm}&=&\frac{\gamma_{k,y}
\cosh\gamma_{k,y}\pm\alpha_{k,y}\sinh{\gamma_{k,y}}}{-\varepsilon
t\sinh\gamma_{k,y}},\\
b_{k,y}^{\pm}&=&\frac{\gamma_{k,y}\exp(\pm\alpha_{k,y})}{-\varepsilon
t\sinh\gamma_{k,y}}.
\end{eqnarray*}

Thus the partition function can be obtained simply by computing the
product of a number of $2\times 2$ matrices. This is a great
simplification to the problem, since the computer time needed for
evaluating this product of $2\times 2$ matrices scales just linearly
with $N_y$. Furthermore, there is no need to store all these
transfer matrices in advance. The computer memory needed in the
calculation is very small. Thus a truly big system with $N_y \sim
10^8$ can be handled without any technical obstacle.

From the partition function, one can readily calculate the free
energy of the system
\begin{equation}
F = -\frac{1}{\beta} \ln Z.
\end{equation}
The magnetization and magnetic susceptibility can then be determined
numerically from the first and second derivatives of the free energy
with respect to the applied magnetic field.

\begin{figure}
\includegraphics[width=0.5\textwidth]{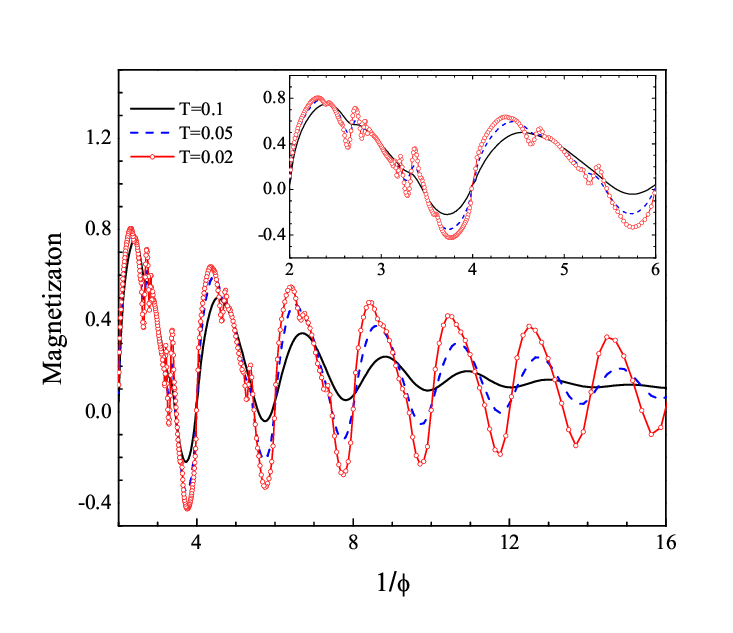}
\caption{ Magnetization of the Hofstadter model at half-filling. The
inset shows more clearly the lattice correction to the dHvA
oscillation in the high field regime. } \label{fig:magneti}
\end{figure}

\begin{figure}[b]
\includegraphics[width=0.5\textwidth]{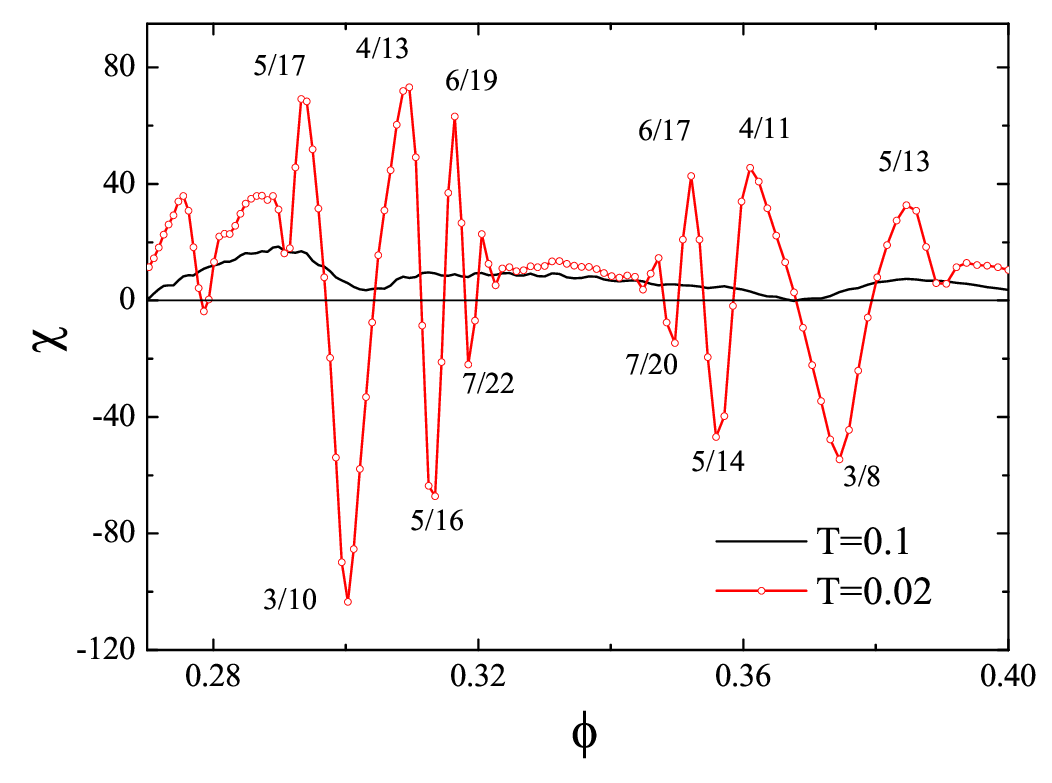}
\caption{\label{fig:magsus} Magnetic susceptibility for the
Hofstadter model at half filling. The values of $\phi$ corresponding
to local maxima and minima in the $T=0.02$ curve are marked. }
\end{figure}

In the Landau gauge, the lattice rotational symmetry is broken. The
finite size effect along the $x$ direction is small. In the
temperature range we have considered, we find that $N_x=50$ is large
enough. However, along the $y$-axis, the finite size effect is
strong. We have evaluated the magnetization at $T=0.02$ by varying
$N_y$ from $500$ to $160000$. We found that the results converge
only after $N_y$ is above $50000$. It indicates that indeed large
lattice systems are needed in order to explore thermodynamic
properties of the Hofstadter model. For higher temperature, the
convergence can be reached with smaller $N_y$. For the results shown
in Figs.~\ref{fig:magneti} and \ref{fig:magsus}, we take $N_x=50$
and $N_y = 80000$ to ensure convergence. For simplicity, here we
consider the half-filling case only. At half filling, the chemical
potential is pinned to $\mu =0$ because of particle-hole symmetry.
In the discussion below, the hopping constant $t$ is set to 1 and
$\varepsilon=0.02$.

Fig.~\ref{fig:magneti} shows the quantum oscillation of
magnetization at three different temperatures. In the low field
limit, the conventional dHvA oscillation is observed. The period of
the oscillation, $\Delta (1/\phi)$, is about 2, consistent with the
result obtained from the formula\cite{solidstatephysics}
\begin{equation}
\triangle ( 1 / \phi )= \frac{ 4\pi^{2}}{S_{F}},
\end{equation}
where $S_F$ is the Fermi volume. At half filling, $S_F = 4\pi^2 /2
= 2\pi^2$. However, with increasing $\phi$, some subtle structures
appear above the conventional dHvA curve expected for two
dimensional electron gas in a magnetic field (see, the inset of
Fig.~\ref{fig:magneti}). These subtle structures become more and
more pronounced with decreasing temperature. They result from the
lattice correction to the energy spectra.

Thermal fluctuation affects strongly on the line shape of
magnetization. At high temperature, say $T=0.1$, the fine fractal
structure of Hofstadter butterfly with energy scale less than $k_BT$
is smeared out by thermal fluctuation. Only the conventional dHvA
oscillation survives, except in the high field limit. However, at
low temperature, say $T = 0.02$, the fine structures of Hofstadter
butterfly with energy scales comparable to $k_BT$ will begin to
influence the magnetic response of the system. It yields the sharp
peaks or deeps observed in the magnetization curve in high fields.
By further reducing temperature, we found that more and more peaks
and deeps, even in the low field range, will emerge from the dHvA
background.

Around each sharp peak or deep, there is a change between
diamagnetism and paramagnetism with increasing temperature. For
example, around $\phi\sim 0.3$, the magnetization decreases with
increasing $\phi$ at $T=0.02$ and the system is diamagnetic;
whereas at $T=0.1$, the magnetization increases with $\phi$ and
the system is paramagnetic. This change from para- to
dia-magnetism is apparently due to the change of energy resolution
since the energy spectrum is unchanged. It is a manifestation of
the fractal structure of Hofstadter butterfly.

Fig.~\ref{fig:magsus} shows the field dependence of magnetic
susceptibility $\chi$ for $\phi$ between 0.27 and 0.4 at $T=0.1$ and
$T=0.02$. At high temperature, $T=0.1$, $\chi$ is paramagnetic.
However, at low temperature, $T=0.02$, $\chi$ oscillates strongly
with $\phi$. It shows a series of local maxima and minima, at which
$\chi$ is positive (paramagnetic) and negative (diamagnetic),
respectively. These extremes appear when the magnetic flux takes
some rational values $\phi = p/q$ (see the rational numbers given in
Fig.~\ref{fig:magsus}). The maxima and minima correspond to odd and
even $q$, respectively.

The appearance of these extremes results clearly from the interplay
between periodic potential and magnetic field. It is strongly
correlated with the density of states of the system at the Fermi
level. The density of states was calculated analytically by Wannier
and co-workers\cite{Wannier}. At half filling, the density of states
vanishes linearly at the Fermi level (namely at a Dirac point) when
$q$ is even\cite{Hasegawa}. However, there is a van Hove singularity
at the Fermi level and the density of states is divergent when $q$
is odd\cite{Hasegawa}. Thus the magnetic response is paramagnetic if
$\phi$ is close to a van-Hove singularity, and diamagnetic if $\phi$
is close to a Dirac point.

However, this connection between the extremes and density of states
seems fragile if considering that there are infinite rational
numbers $p/q$ with even \emph{and} odd denominators in an arbitrary
small but finite interval of $\phi$. In other words, near any
rational number, say $\phi=4/13$, there are infinite other $\phi$ at
which the density of states at the Fermi level can be either zero or
divergent. So how can we attribute the orbital paramagnetism at
$4/13$ to the van-Hove singularity in the density of states?

This problem can be resolved by considering the hierarchical
structure of Hofstadter butterfly and the temperature smearing of
the band structure. At the first rank of hierarchy, the Hofstadter
butterfly is divided into several subcells\cite{Hofstadter}. These
subcells can be further divided recursively into many sub-subcells.
This hierarchical recursion defines a parallel iterative
transformation. After this transformation, any rational $\phi$ can
be finally reduced to a simple rational number, which is equal to
either $1/q'$ or $1-1/q'$, where $q'$ is an integer.

For example, $\phi = 4/13$ can be reduced to $4/5$ after only one
iteration. This means in the first order subcell centered at the
Fermi level, there are five subbands and the middle one has a
divergent density of states crossing the Fermi level. On the other
hand, $\phi=401/1300$, which is a value very close to $4/13$, can be
reduced to $3/4$ after 19 iterations. This means that, in the 19th
order subcell, there are four subbands and the middle two meet at
the Fermi level. In this case, the Fermi level is a Dirac point and
the corresponding density of states vanishes. However, the
characteristic energy scale of the 19th order subcell is very small
compared with the thermal energy $k_B T$ at $T=0.02$. Therefore, the
contribution by the singularity at $\phi=401/1300$ to $\chi$ is
completely smeared out by thermal fluctuation and only the peak at
$\phi = 4/13$ can be seen at $T=0.02$.

This can be seen more clearly by integrating out the density of
states in an interval of $k_BT$ around the Fermi level. We find that
the integral at $\phi=401/1300$ is hardly different from that at
$\phi=4/13$. Therefore, around $\phi \sim 4/13$ the density of state
at the Fermi level is determined by the van-Hove singularity in the
first subcell at $4/13$. Similar argument can be applied to $\phi$
with even $q$. The difference is that in that case the density of
states is dominated by the Dirac points.

The above argument implies that the higher order subcells of
Hofstadter butterfly can be probed by increasing the energy
resolution. Thus more van-Hove singularities and Dirac points can be
discerned by lowering temperature. However, in the limit of zero
temperature, the susceptibility is no longer a well defined
quantity, since it can oscillate between paramagnetism and
diamagnetism in an infinitesimally small interval of $\phi$. This is
consistent with the fact that that ground state energy is not
differentiable with respect to $\phi$.

In conclusion, we have introduced a quantum transfer matrix method
to study thermodynamic properties of the Hofstadter model on square
lattices. This method allows thermodynamic quantities to be
accurately and efficiently evaluated without suffering from the
finite size effect. Our study suggests that the Hofstadter butterfly
can be probed by thermodynamic measurements. In particular, the
magnetic susceptibility is sensitive to the change of the density of
states. It shows a paramagnetic peak if the density of states has a
van Hose singularity at the Fermi level or a diamagnetic deep if the
Fermi surface is a Dirac point. Thus the measurement of magnetic
susceptibility, which can be materialized on the
superlattice\cite{Albrecht} or cold atom\cite{Umucalilar} systems,
can reveal not only the fractal structure of spectra, but also the
density of states of the Hofstadter model.

Acknowledgement: We thank Prof. L. Yu for useful discussions. This
work was supported by the NSF-China and the National Program for
Basic Research of MOST, China.

\end{document}